\documentclass[10pt,aps,amsmath,amssymb,superscriptaddress,nobalancelastpage,prl,twocolumn]{revtex4-2}
\usepackage{ulem}
\usepackage[utf8]{inputenc}
\usepackage{textcomp}
\DeclareUnicodeCharacter{2212}{\ensuremath{-}}

\usepackage{xr-hyper}
\usepackage{bm}
\usepackage{braket}
\usepackage{diagbox}
\usepackage{graphicx}
\usepackage{hyperref}
\hypersetup{colorlinks,linkcolor=blue,urlcolor=blue,citecolor=blue}
\usepackage{siunitx}
\usepackage{varioref}
\usepackage{xcolor}
\usepackage{tabularx}
\begin{document}
\title{Magnetic Excitations of a Half-Filled Tl-based Cuprate}

\author{I.~Bia\l{}o}
\email{izabela.bialo@physik.uzh.ch}
\affiliation{Physik-Institut, Universit\"{a}t Z\"{u}rich, Winterthurerstrasse 
190, CH-8057 Z\"{u}rich, Switzerland}

\author{Q.~Wang}
\email{qwang@cuhk.edu.hk}
\affiliation{Department of Physics, The Chinese University of Hong Kong, Shatin, Hong Kong, China}
\affiliation{State Key Laboratory of Quantum Information Technologies and Materials, The Chinese University of Hong Kong, Shatin, Hong Kong, China}

\author{J.~Küspert}
\affiliation{Physik-Institut, Universit\"{a}t Z\"{u}rich, Winterthurerstrasse 190, CH-8057 Z\"{u}rich, Switzerland}
\affiliation{European Synchrotron Radiation Facility, 71 Avenue des Martyrs, 38043 Grenoble, France}

\author{X.~Hong}
\affiliation{Physik-Institut, Universit\"{a}t Z\"{u}rich, Winterthurerstrasse 190, CH-8057 Z\"{u}rich, Switzerland}
\affiliation{Department of Physics, The Chinese University of Hong Kong, Shatin, Hong Kong, China}

\author{L.~Martinelli}
\affiliation{Physik-Institut, Universit\"{a}t Z\"{u}rich, Winterthurerstrasse 190, CH-8057 Z\"{u}rich, Switzerland}

\author{O.~Gerguri}
\affiliation{Physik-Institut, Universit\"{a}t Z\"{u}rich, Winterthurerstrasse 190, CH-8057 Z\"{u}rich, Switzerland}
\affiliation{PSI Center for Neutron and Muon Sciences CNM, 5232 Villigen PSI, Switzerland}

\author{Y.~Chan}
\affiliation{Department of Physics, The Chinese University of Hong Kong, Shatin, Hong Kong, China}

\author{K.~von~Arx}
\affiliation{Physik-Institut, Universit\"{a}t Z\"{u}rich, Winterthurerstrasse 190, CH-8057 Z\"{u}rich, Switzerland}
\affiliation{Department of Physics, Chalmers University of Technology, SE-412 96 G\"{o}teborg, Sweden}

\author{O.~K.~Forslund}
\affiliation{Physik-Institut, Universit\"{a}t Z\"{u}rich, Winterthurerstrasse 190, CH-8057 Z\"{u}rich, Switzerland}
\affiliation{Department of Physics and Astronomy, Uppsala University, Box 516, SE-75120 Uppsala, Sweden}

\author{W.~R.~Pude\l{}ko}
\affiliation{Physik-Institut, Universit\"{a}t Z\"{u}rich, Winterthurerstrasse 190, CH-8057 Z\"{u}rich, Switzerland}
\affiliation{Swiss Light Source, Paul Scherrer Institut, CH-5232 Villigen PSI, Switzerland}

\author{C.~Lin}
\affiliation{Physik-Institut, Universit\"{a}t Z\"{u}rich, Winterthurerstrasse 190, CH-8057 Z\"{u}rich, Switzerland}
\affiliation{Stanford Synchrotron Radiation Lightsource, SLAC National Accelerator Laboratory, Menlo Park, USA}

\author{N.~C.~Plumb}
\affiliation{Swiss Light Source, Paul Scherrer Institut, CH-5232 Villigen PSI, Switzerland}

\author{Y.~Sassa}
\affiliation{Department of Physics, Chalmers University of Technology, SE-412 96 G\"{o}teborg, Sweden}

\author{D.~Betto}
\affiliation{European Synchrotron Radiation Facility, 71 Avenue des Martyrs, 38043 Grenoble, France}

\author{N.~B.~Brookes}
\affiliation{European Synchrotron Radiation Facility, 71 Avenue des Martyrs, 38043 Grenoble, France}

\author{M.~Rosmus}
\affiliation{Université Paris-Saclay, CNRS, Institut des Sciences Moléculaires d'Orsay, 91405, Orsay,France}
\affiliation{Solaris National Synchrotron Radiation Centre, Jagiellonian University, Czerwone Maki 98, 30-392 Krak\'{o}w, Poland}

\author{N.~Olszowska}
\affiliation{Solaris National Synchrotron Radiation Centre, Jagiellonian University, Czerwone Maki 98, 30-392 Krak\'{o}w, Poland}

\author{M.~D.~Watson}
\affiliation{Diamond Light Source, Harwell Campus, Didcot, Oxfordshire OX11 0DE, United Kingdom}

\author{T.~K.~Kim}
\affiliation{Diamond Light Source, Harwell Campus, Didcot, Oxfordshire OX11 0DE, United Kingdom}

\author{C.~Cacho}
\affiliation{Diamond Light Source, Harwell Campus, Didcot, Oxfordshire OX11 0DE, United Kingdom}

\author{M.~Horio}
\affiliation{Institute for Solid State Physics, The University of Tokyo, Kashiwa, Chiba 277-8581, Japan}

\author{M.~Ishikado}
\affiliation{Comprehensive Research Organization for Science and Society (CROSS), Tokai, Ibaraki 319-1106, Japan}

\author{H.~M.~Rønnow}
\affiliation{Laboratory for Quantum Magnetism, École Polytechnique F\'{e}d\'{e}rale de Lausanne (EPFL), CH-1015 Lausanne, Switzerland}

\author{J.~Chang}
%\email{johan.chang@physik.uzh.ch}
\affiliation{Physik-Institut, Universit\"{a}t Z\"{u}rich, Winterthurerstrasse 190, CH-8057 Z\"{u}rich, Switzerland}

%\date{\today}

\begin{abstract}
Strong electron correlations drive Mott insulator transitions. Yet, there exists no framework to classify Mott insulators by their degree of correlation. Cuprate superconductors, with their tunable doping and rich phase diagrams, offer a unique platform to investigate the evolution of these interactions. However, spectroscopic access to a clean half-filled Mott-insulating state is lacking in compounds with the highest superconducting onset temperature. To fill this gap, we introduce a pristine, half-filled thallium-based cuprate system, Tl$_2$Ba$_5$Cu$_4$O$_{x}$. Using high-resolution resonant inelastic x-ray scattering, we probe long-lived magnon excitations and uncover a pronounced kink in the magnon dispersion, marked by a simultaneous change in group velocity and lifetime broadening. Modeling the dispersion within a Hubbard–Heisenberg approach, we extract the interaction strength and compare it with other cuprate systems. Our results establish a cuprate universal relation between electron-electron interaction and magnon zone-boundary dispersion. Superconductivity seems to be optimal at intermediate correlation strength, suggesting an optimal balance between localization and itinerancy. 
\end{abstract}
\maketitle

\section{Introduction}
\label{Introduction}

Electron-electron interactions mediate a wealth of emergent phenomena such as magnetism, Mott insulating behavior, and unconventional superconductivity~\cite{dagotto_complexity_2005}. In Fermi liquids, the electronic mass renormalization is a gauge of electron-electron correlations, and the Kadowaki–Woods ratio yields an universal relation between mass and scattering time~\cite{kadowaki_universal_1986,jacko_unified_2009}. Interestingly, for Mott insulators no such analogous ratio exists. Establishing such a quantity could provide a powerful framework for uncovering links between correlation strength and emergent phases, including superconductivity~\cite{jiang_superconductivity_2019}. As magnetic interactions provide a direct fingerprint of electronic correlations, exploring magnetic excitations offers a natural route to quantify interaction strength~\cite{coldea_spin_2001,braicovich_dispersion_2009}. Cuprates, as archetypal Mott systems evolving into Fermi liquids with tunable doping, provide a unique opportunity to link magnetic excitation spectra to the underlying correlations and ultimately to the superconducting transition temperature $T_c$~\cite{lee_doping_2006,keimer_quantum_2015}.

\begin{figure*}
\centering
\includegraphics[scale=1]{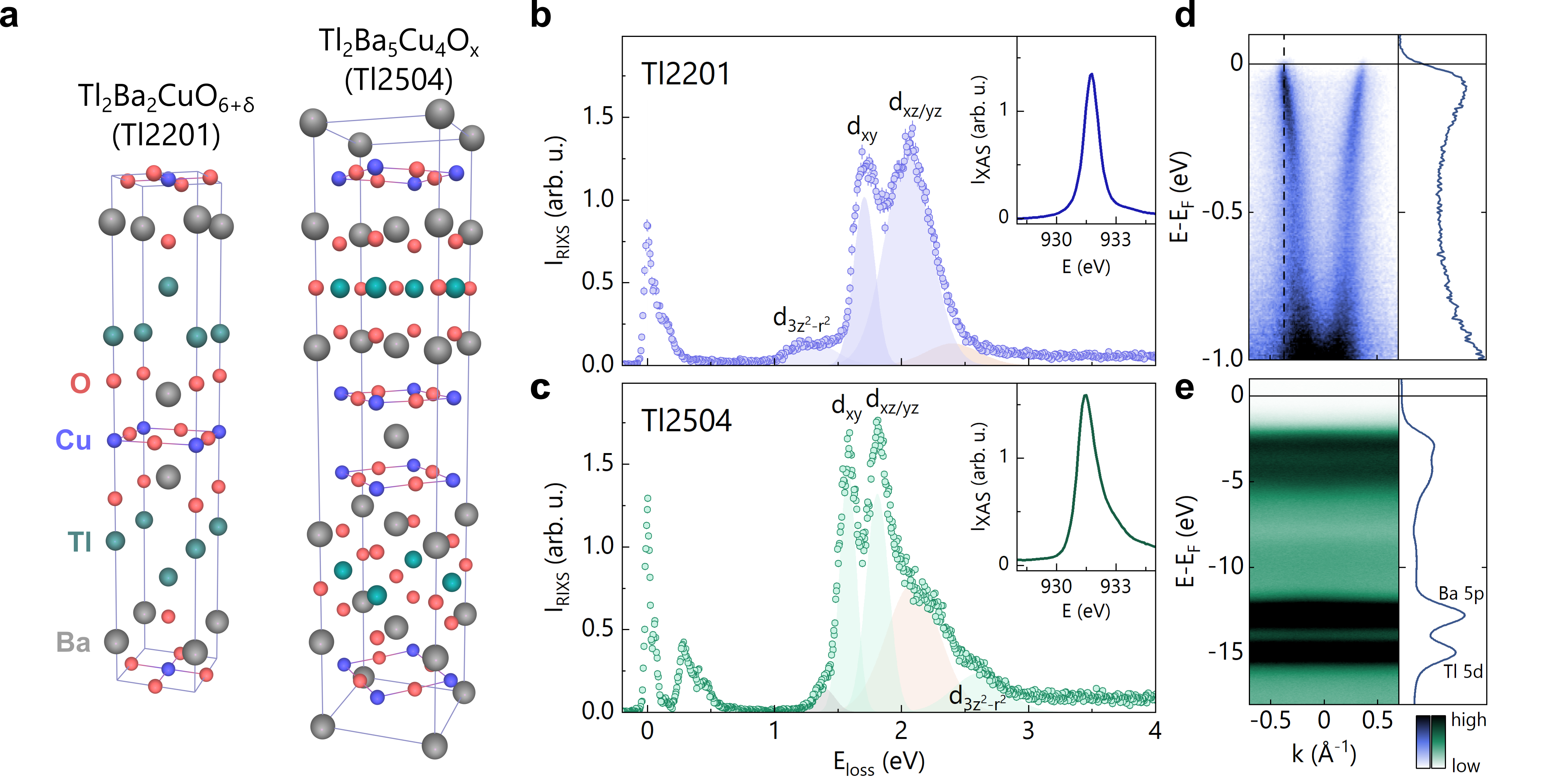}
\caption{\textbf{Characteristic of single and double-layer Tl-based cuprates} (a) Crystal structures of Tl2201 and Tl2504 (Cu-O bonds are indicated). (b,c) Low temperature RIXS spectra at $(h,0)=(0.38,0)$ measured with $\pi$ polarized incident light. Main orbital excitations are determined from fitting with Gaussians and are denoted by shaded areas. XAS profiles of Cu $L_3$ absorption edges, which correspond to the transitions from \textit{2p} to empty \textit{3d} states, are presented as insets. RIXS spectra were collected with incident light energy tuned to the maximum of the Cu $L_3$ edge. The RIXS spectrum for $p=0.25$ doped Tl2201 is adapted from~\cite{Tam2022}. (d,e) ARPES band maps taken along the nodal direction for metallic, overdoped Tl2201 ($p=0.25$)~\cite{kramerPRB2019} and insulating Tl2504, respectively. The corresponding energy distribution curves in the right-hand panels were integrated within (d) $k_F-0.05>k>k_F+0.05$ with $k_F$ marked by a dashed line or (e) within the presented range of \textit{k}.}
\label{Fig1}
\end{figure*}

The cuprate phase diagram is pieced together by combining insights from multiple compounds with distinct chemical compositions. Systems such as La$_{2-x}$Sr$_x$CuO$_4$ cover the entire doping range, but are reached by chemical substitution, which inevitably introduces disorder~\cite{Eisaki2004,Alloul2009}. 
Therefore, more attention has been given to structurally simpler and more pristine compounds with higher $T_c$'s  such as Tl$_2$Ba$_2$CuO$_{6+\delta}$ (Tl2201)~\cite{VignolleNature2008,plate_fermi_2005, Tam2022} or HgBa$_2$CuO$_{4+\delta}$~\cite{barisic_universal_2013,yu_unusual_2020}. However, these systems cannot be synthesized in the highly underdoped regime. Within materials with largest $T_c$, YBa$_2$Cu$_3$O$_{6+\delta}$ stands out as an exception. Yet, its resonant inelastic x-ray scattering (RIXS) spectra differ significantly from those of simpler cuprate insulators~\cite{minola_collective_2015} i.e. La$_2$CuO$_4$ (LCO)~\cite{headings_anomalous_2010} and CaCuO$_2$ (CCO)~\cite{martinelli_decoupling_2025}, primarily due to structural complexities including Cu–O chains, bilayer modulations, and oxygen disorder. As a consequence, none of the clean high-$T_c$ systems provide high-quality spectroscopic data on the Mott-insulating state. A structurally simple compound that combines high $T_c$ with clear access to half-filling is therefore still missing. 

Here, we present a half-filled version of a thallium-based cuprate, Tl$_2$Ba$_5$Cu$_4$O$_{x}$ (Tl2504), and introduce its spectroscopic characteristics, and compare with the overdoped, metallic Tl2201. In particular, we focus on high-resolution RIXS measurements to probe its magnetic excitation. Our experiments reveal a large zone-boundary dispersion accompanied by a sudden velocity change -- a pronounced kink in the magnon dispersion -- concomitant with a decrease in the magnon lifetime. We analyze our data within a Heisenberg-Hubbard model, introducing a momentum-dependent renormalization factor~\cite{wang_magnon_2024}. This enables us to identify the interaction strength of Tl2504 and make direct comparison to other cuprate systems. 

\section{Results}
\label{Results}

\textit{Crystal-Field Environment --} Fig.~1(a) illustrates crystal structures of Tl2201 and Tl2504~\cite{Hasegawa1996}, which exhibit distinct $c$-axis lattice parameters and differ in the orientation of their CuO$_2$ planes relative to the Ba lattice (rotated by 45$^{\circ}$). These distinctions manifest themselves in the diffraction characteristics (see Supplementary Information (SI), Fig.~S1). They are also revealed in the structure of $dd$ excitations present in the high-energy part of the RIXS spectra (Fig.~\ref{Fig1}(b,c)). \textit{dd}-excitations in cuprates are composed of three main features (green/purple peaks) corresponding to transitions between the $d_{x^2-y^2}$ ground state and the other $3d$ orbitals split by the tetragonal crystal field. The splitting of $t_{2g}$ orbitals ($d_{xy}$ and the degenerate $d_{xz/yz}$) is smaller in the case of the bilayer system, indicating a lower buckling of CuO$_2$ planes expected for multi-layered systems. The splitting of $e_g$ states (between $d_{x^2-y^2}$ and $d_{z^2}$) scales with the distance between CuO$_2$ planes and apical ligands~\cite{moretti_sala_energy_2011}, and is greater for Tl2504 than for Tl2201. The \textit{dd} profiles exhibit additional spectral weight at $2-2.5$~eV (brown peak) that can be attributed to oxygen vacancies~\cite{moretti_sala_energy_2011} and was previously observed in (Sr/Ca)CuO$_2$ cuprates~\cite{martinelli_collective_2024}. In Tl2504, the preceding peak at 1.4~eV (gray) can be assigned to a hybridization of Cu $3d$ orbitals with orbitals of another atom in a structure, analogously to the hybridization of Ni $3d$ and rare-earth  $5d$ orbitals in nickelate oxides~\cite{yan_persistent_2025,gao_magnetic_2024}.

\begin{figure*}
\centering
\includegraphics[scale=1]{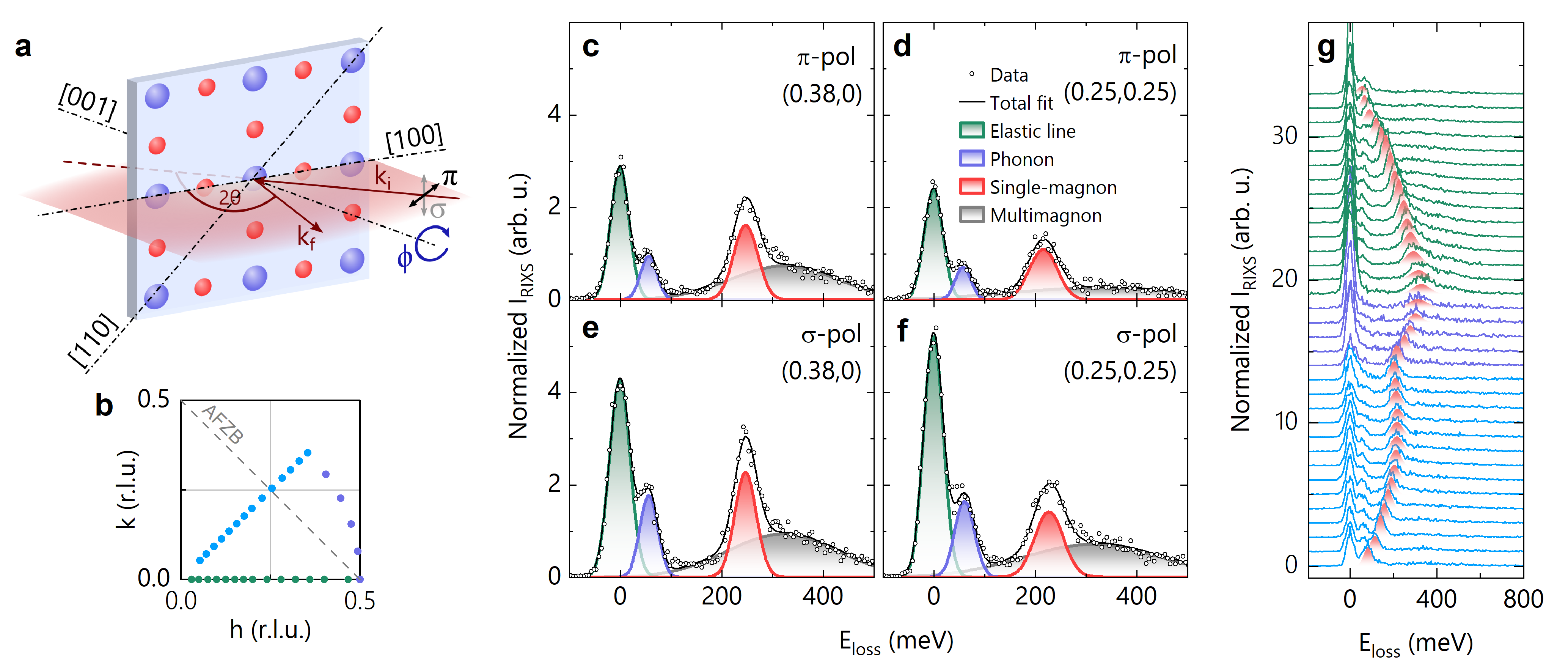}
\caption{\textbf{RIXS studies of undoped Tl2504} (a) Scattering geometry used for the RIXS experiment. $k_i$ and $k_f$ represent the momentum of the incident and scattered photons within the scattering plane (red shadow) perpendicular to the CuO$_2$ plane of the sample. The incident beam is polarized ($\pi$ or $\sigma$). (b) Reciprocal space probed by RIXS measurements. The gray dashed line represent the antiferromagnetic zone boundary (AFZB). (c-f) The low-energy region of normalized RIXS spectra fitted by four Gaussian components. A solid black line represents the sum of the fitting components. (g) Normalized RIXS spectra for the $\pi$-polarized beam recorded along three momentum trajectories defined by the colors in panel (b). The magnon contribution to RIXS spectra, based on fitting results, is indicated as a red shadow.}
\label{Fig2}
\end{figure*}

\textit{Mott-Insulating State --} By varying the oxygen content ($\delta$), Tl2201 bridges the optimal to overdoped range of doping (\textit{p}) within the cuprate phase diagram~\cite{keimer_quantum_2015}. 
%split in to two , doping separately 
The RIXS spectrum shown in Fig.~\ref{Fig1}b is recorded on an overdoped hole concentration $p=0.25$. To illustrate the metallic nature of this doping region, we show (Fig.~\ref{Fig1}d) an angle-resolved photoemission spectroscopic (ARPES) spectrum with bands crossing the Fermi level at nodal points of the Brillouin zone (corresponding magnetization data are presented in SI, Fig.~S2). ARPES spectra recorded on Tl2504 (Fig.~\ref{Fig1}e) display no evidence of low-energy quasiparticles~\cite{plate_fermi_2005,VignolleNature2008}. Instead, the spectra display a momentum-independent $\sim1.8$~eV electronic gap. This is comparable to what is observed in the other Mott insulating cuprates~\cite{Hashimoto2014}, suggesting that Tl2504 crystals represent the first reported realization of a half-filled Tl-based cuprate.

\textit{Low-energy Excitations --} Additional indication of the insulating nature of Tl2504 is provided by the low-energy region of the RIXS spectra (below 0.5 eV) (Fig.~\ref{Fig1}c). Here, a nearly resolution-limited sharp excitation is visible, which can be identified as a magnon mode. The observed long magnon lifetime is a characteristic that is typically observed only in the half-filled Mott insulating state~\cite{peng_influence_2017, dean_persistence_2013}. In contrast, magnetic excitations in the metallic Tl2201 (at 0.3~eV), measured at the same momentum transfer, exhibit significant damping in intensity and energy (Fig.~\ref{Fig1}b), consistent with previous reports on optimally and overdoped cuprates~\cite{dean_persistence_2013, TaconPRB2013}.

A detailed analysis of the low-energy part of the Tl2504 RIXS spectra is shown in Fig.~\ref{Fig2}. The spectra were acquired using the $\pi$ and $\sigma$ polarizations of incident photons (see experimental geometry in Fig.~\ref{Fig2}a), covering the part of the Brillouin zone presented in Fig.~\ref{Fig2}b. The observed intensity can be analyzed with a four-component model (Fig.~\ref{Fig2}(c-f)). The sharpness of the reported excitations enables the use of a Gaussian profile for each of them. The excitation at $\sim60$~meV matches the CuO bond-stretching phonon mode previously observed in Cu \textit{L}-edge studies of cuprates~\cite{Tam2022,chaix_dispersive_2017}. The single (and multi) magnon excitations are visible for both light polarizations. The single magnon contributions to the experimental spectra, obtained through the aforementioned fitting procedure, are marked in Fig.~\ref{Fig2}g by a red-shadowed area.

\begin{figure*}
\centering
\includegraphics[width=1\textwidth]{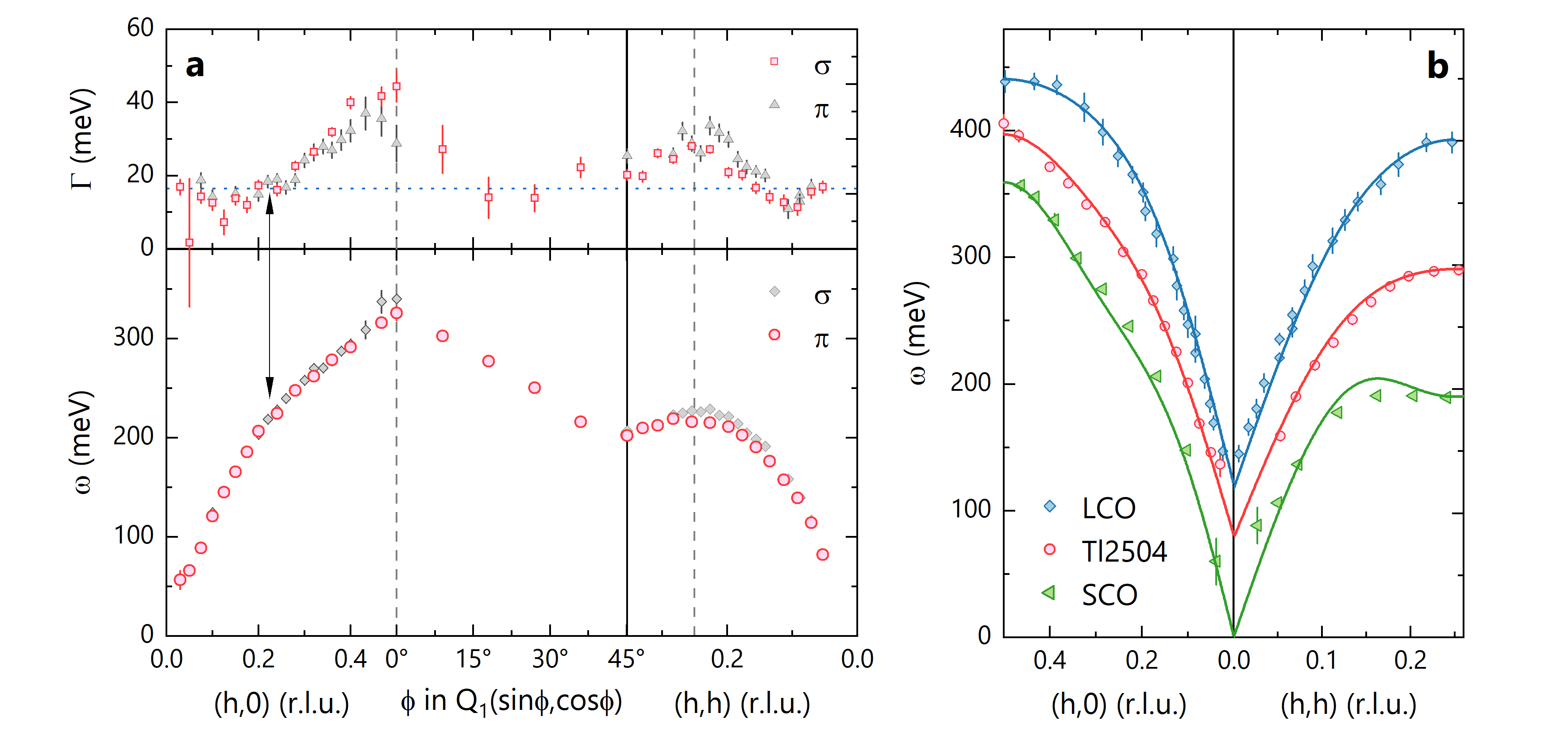}
\caption{\textbf{Single-magnon dispersion in half-filled Tl2504} (a) Single-magnon dispersion and its inverse lifetime extracted from RIXS measurements performed with $\sigma$ and $\pi$ polarized light. The black arrow indicates simultaneous change of magnon energy and inverse lifetime. The gray dashed line represents the AFZB, while the azimuthal part of the dispersion is defined by $Q_1=0.5$. A blue dotted line represent the experimental energy resolution. (b) Comparison of magnon dispersion along $(h,0)$ and $(h,h)$ directions for selected cuprate systems. Solid lines are corresponding fits using the Hubbard model including renormalization factor (see text). Datasets for Tl2504 and LCO are shifted by a constant in energy scale, correspondingly by $0.8$~eV and $0.12$~eV. Data for LCO (SCO) has been adapted from~\cite{headings_anomalous_2010} (\cite{wang_magnon_2024}).}
\label{Fig3}
\end{figure*}

\textit{Single-magnon dispersion --}
The magnon dispersion of Tl2504 is shown in Fig.~\ref{Fig3}a. It is characterized by a zone-boundary dispersion $E_{\mathrm{ZB}}$, between (0.5,0) and (0.25,0.25) points, of about 100 meV, indicating significant ring- and/or higher-order magnetic exchange interactions~\cite{coldea_spin_2001,martinelli_fractional_2022}. The lack of a pronounced gap around the zone center implies that the bilayer coupling~\cite{reznik_direct_1996,hayden_high-frequency_1996} is below our resolving power. As in YBa$_2$Cu$_3$O$_{7-x}$~\cite{tacon_intense_2011} and NdBa$_2$Cu$_3$O$_{7-x}$\cite{peng_influence_2017}, our RIXS data on Tl2504 deos not resolve simultaneously optical and acoustic magnon branches. Across the Brillouin zone, magnon excitations remain resolution-limited, except near the AFZB, where they broaden and their inverse lifetime increases. A similar suppression of the magnon spectral weight around the $(0.5,0)$ point has previously been reported in several cuprate systems~\cite{headings_anomalous_2010,martinelli_fractional_2022,christensen_quantum_2007}, as well as in other square-lattice systems with spin $1/2$ \cite{christensen_quantum_2007,gretarsson_persistent_2016}. A key new observation reported here is that the magnon broadening along the $(h,0)$ direction correlates with a clear change in the slope of the magnon dispersion (indicated by a black arrow). This concomitant change of magnon lifetime and velocity suggests an interaction-induced effect. 

Although not discussed previously in literature, the change in slope of the magnon dispersion around the middle of the zone is also observed in other cuprate systems, such as SrCuO$_2$ (SCO)~\cite{wang_magnon_2024} (green points in Fig.~\ref{Fig3}b) or CaCuO$_2$~\cite{martinelli_fractional_2022}. It should be underlined, however, that this behavior is not generic to all cuprates. In systems with very strong correlations ($U/t\gg8$), the magnon dispersion is following a $\sin(k)$-shaped momentum dependence~\cite{coldea_spin_2001,headings_anomalous_2010}. A canonical example is LCO, which dispersion is shown in blue in Fig.~\ref{Fig3}b.

\section{Discussion}
\label{Discussion}
In the cuprates, the Mott insulating state around half-filling is characterized by an electronic excitation gap and well-defined magnon excitations. As we observe a large quasiparticle gap with ARPES and resolution limited magnon excitations with RIXS, we conclude our Tl2504 crystal is Mott insulating at (near) half-filling. Magnons on a square lattice with nearest-neighbor interaction (only) produce an isotropic dispersion around the zone center. As such, the pronounced zone boundary dispersion, observed in Tl2504 and other cuprates, suggests that higher-order magnetic exchange couplings are significant and indicate the presence of substantial four-body magnon interactions~\cite{ColdeaPRL01}. These can be accounted for by the Hubbard model projected into a Heisenberg Hamiltonian, where the Coulomb interaction $U$, nearest, next- and next-next nearest neighbor hopping integrals ($t,t^\prime$ and $t^{\prime\prime}$) are included. The ring-exchange interaction $J_c\sim t^4/U^3$ introduces a zone boundary dispersion. Likewise, higher-order hopping terms ($t^\prime$, and $t^{\prime\prime}$) produce four-body interactions that contribute to enhancing the zone-boundary dispersion~\cite{delannoy_low-energy_2009,ivashko_damped_2017,dalla_piazza_unified_2012}. 

To discuss the magnon dispersion observed in Tl2504 (and other cuprates), we use an approach based on non-local, single-magnon interactions. The Hubbard-Heisenberg model dictates $\hbar\omega_k= Z_c(k)\epsilon_k$ where $\epsilon_k$ is the bare magnon dispersion and $Z_c(k)=Z_c^0(1+f_k)$ is a non-local renormalization factor defined by a momentum-dependent function $f_k$~\cite{wang_magnon_2024}. Including the $k$‑dependence of the $Z$ factor is necessary to reproduce the discontinuities in magnon velocity, as well as the substantial renormalization of magnon energy at the AFZB, around the $(0.25,\,0.25)$ point~\cite{wang_magnon_2024}. When $U/t\rightarrow\infty$, as represented by LCO, $Z$ factor becomes nearly momentum-independent. The fitted magnon dispersions obtained from this model are presented in Fig.~\ref{Fig3}b by solid lines (a detailed description of the fitting procedure is included in the Methods section). Enhanced contributions from higher-order exchange interactions are reflected in finite values of $f_k$, giving rise to pronounced magnon dispersion at the Brillouin zone boundary and anomalous magnon velocity effects. The strength of such higher-order exchange terms grows as the electronic correlation parameter $U/t$ decreases. In the cuprates, $U/t$ has been shown to diminish with increasing apical oxygen distance from the CuO$_2$ planes.

Fitting parameters extracted from the Hubbard model allow us to place Tl2504 within the broader context of cuprate materials. Based on the key energy scale in correlated systems, reflected by the $U/t$ ratio, Tl2504 is classified as a moderately correlated compound with $U/t = 6.91$, in contrast to more strongly correlated systems such as LCO, where $U/t = 9.79$ is obtained.

\begin{figure}
\centering
\includegraphics{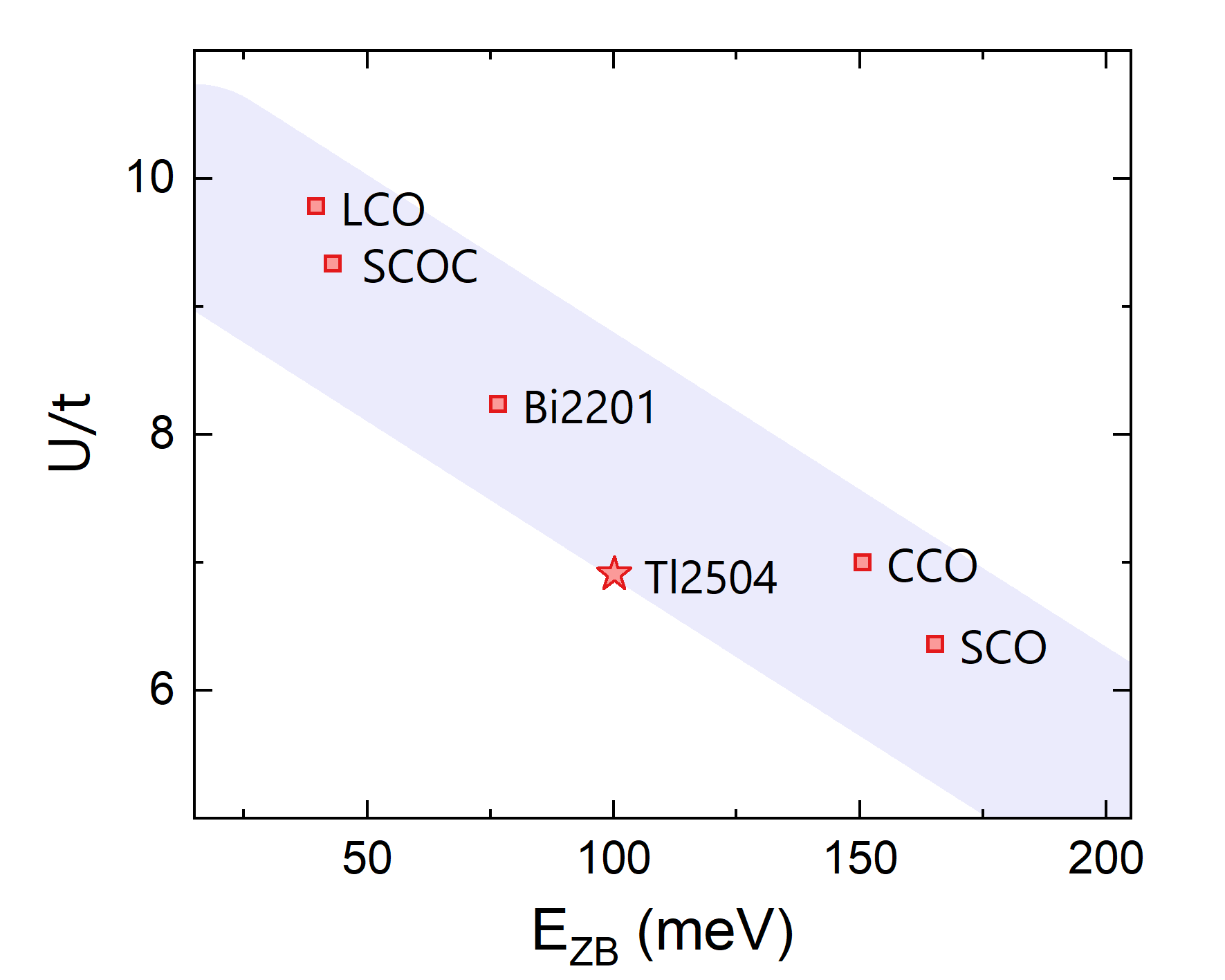}
\caption{\textbf{Strength of electronic correlations} From parametrization of magnon dispersions with a Hubbard model (see text), $U/t$ and $E_{\mathrm{ZB}}$ are extracted for various cuprate compounds. Fitting of magnon dispersions used data from Tl2504 (this work, star), LCO~\cite{headings_anomalous_2010}, Sr$_2$CuO$_2$Cl$_2$ (SCOC)~\cite{plumb_high-energy_2014}, Bi$_2$Sr$_{0.9}$La$_{1.1}$CuO$_6$ (Bi2201)~\cite{peng_influence_2017}, CCO~\cite{martinelli_fractional_2022} and SCO~\cite{wang_magnon_2024}.}
\label{Fig4}
\end{figure}

Figure~\ref{Fig4} presents the correlation strength $U/t$ alongside the zone-boundary magnon energy $E_{\mathrm{ZB}}$, which reflects the influence of four-body interactions~\cite{delannoy_low-energy_2009,ivashko_damped_2017,dalla_piazza_unified_2012}. Remarkably, the ratio of these two energy scales is nearly constant across a wide range of cuprates. Recent studies have shown that $E_{\mathrm{ZB}}$ correlates with the energy of the Cu $3d_{z^2}$ orbital and with the distance to the apical oxygen~\cite{peng_influence_2017,moretti_sala_energy_2011}, indicating that the apical oxygen distance primarily controls the in-plane correlation strength. This suggests that realistic models of cuprates include also out-of-plane orbitals~\cite{weber_apical_2010,sakakibara_two-orbital_2010}. Indeed, three-orbital Hubbard models have successfully described the magnon physics of cuprates~\cite{wang_magnon_2018,kung_characterizing_2016,spalek_superconductivity_2022}.

\section{Conclusions}
\label{Conclusions}
In this work, we have successfully synthesized a pristine Tl-based, half-filled cuprate material, Tl2504. We prove its insulating properties and characterize its crystal structure. High‑resolution Cu $L_3$‑edge RIXS measurements reveal a large zone‑boundary magnon dispersion, and a pronounced kink in the dispersion accompanied by magnon lifetime broadening, directly signaling the presence of significant four‑body exchange interactions. By fitting the dispersion to a Hubbard–Heisenberg model with a momentum‑dependent renormalization factor $Z_c(\mathbf{k})$, we extract a moderate correlation strength. Moreover, we show that the ratio of the zone‑boundary energy $E_{\rm ZB}$ to $U/t$ collapses onto a universal trend across a broad range of cuprates, underscoring the central role of apical‑oxygen geometry and interlayer screening in tuning the in‑plane spin dynamics. 
In summary, our results indicate that high-temperature superconductivity may emerge most favorably at an optimal, intermediate level of electronic correlation, striking a delicate balance between electron localization and itinerancy.

\section{Methods}
\label{Experimental}
\noindent\textbf{Sample growth and characteristic:}
Tl2504 single crystals were grown using a self-flux technique. Two $0.5\times0.5\times0.1$~mm samples were measured. \textit{c}-axis crystal lattice parameter, characteristic for the bilayered system~\cite{Hasegawa1996}, was determined as $27.10~\text{\AA}$ based on X-ray diffraction studies of $(002)$ \textit{out-of-plane} Bragg reflections performed at the ID32 beamline at the European Synchrotron Radiation Facility (ESRF) (see Fig.~S1 in SI). The in-plane parameters of Tl2504 are almost $\sqrt{2}$ times the values reported for the single-layer compound~\cite{Hasegawa1996}, with $a=b=5.5$~\AA. \\

\noindent\textbf{RIXS experiments:}
Data for Tl2504 and NCO were collected at the ID32 beamline at the ESRF~\cite{brookes_beamline_2018}. Spectra were recorded in the medium resolution instrument configuration ($\gamma\approx33$~meV) and with incident light polarization of both linear vertical ($\sigma$) and linear horizontal ($\pi$). $\gamma$ was determined as the full-width-at-half-maximum of the elastic signal from silver paint. The incident photon energy was tuned to the absorption peak of the Cu $L_3$ resonance edge ($\sim 931$~eV). The orientation in reciprocal space was determined by Laue diffraction (\textit{in-plane} components) and on the base of $(002)$ \textit{out-of-plane} Bragg reflections. The wave vector $\textbf{Q}$ in $(q_x, q_y, q_z)$ is defined as ($h, k, \ell$) = $(q_x a_{\mathrm{T}}/2\pi, q_y a_{\mathrm{T}}/2\pi, q_z c_{\mathrm{T}}/2\pi)$ reciprocal lattice units (r.l.u.). We adopt a tetragonal reference unit cell with $a_{\mathrm{T}} = a/\sqrt{2}$ and $c_{\mathrm{T}} = c$, rotated by 45$^\circ$ along $c$-axis with respect to the original unit cell (Fig.~\ref{Fig1}a). In this way, $a_{\mathrm{T}}$ lies along the Cu--O--Cu bond directions in the CuO$_2$ planes and simplifies the description of magnetic excitations compared to other cuprates. The $(h,k)$ plane was scanned by changing the orientation of the sample in $\theta$ and using a fixed scattering angle $2\theta = 149.5^{\circ}$. All data were collected at $20$~K, under ultra-high vacuum conditions ($10^{-10}$~mbar). 

The elastic peaks were fitted using a Gaussian function to determine the zero-energy loss reference. All spectra were normalized in intensity to the area within the $1<E<3$~eV energy range. The error bars for the intensity of the RIXS (Fig.~\ref{Fig1}) are calculated as a square root of the total photon count, whereas the error bars for the fitting parameters correspond to three standard deviations (Fig.~\ref{Fig3}a). The inverse lifetime $\Gamma$ is defined by the energy width of the single-magnon excitation $\mathcal{G}$ and the corresponding experimental resolution $\gamma$, according to the formula $\Gamma = \frac{1}{2} \sqrt{\mathcal{G}^2 - \gamma^2}$. \\

\noindent\textbf{ARPES experiments:}
Measurements of Tl2504 single crystals were performed at the I05 and URANOS beamlines at the Diamond Light Source in the United Kingdom and the SOLARIS National Synchrotron Radiation Center in Poland, respectively. The samples were electrically grounded using silver epoxy and equipped with a top post for cleaving. For further strengthening of the electrical connectivity to the cleaved surface, a graphite spray was used. These measures eliminated the charging effects. The samples were cleaved in situ at $200$~K with vacuum conditions better than \num{2d-10}~mbar. The Fermi level was calibrated by reference to the electrically connected gold foil. The measurement conditions (including those for Tl2201) are summarized in Table~\ref{tab:ARPES}.

\begin{table}[h]
\centering
\caption{\textbf{Experimental parameters for ARPES measurements} Comparison of measurement conditions at the different synchrotron beamlines,  for different Tl-based cuprates. $E_{res}$ stands for a value of energy resolution of the corresponding experiment.}
\begin{ruledtabular}
\begin{tabular}{llll}
%\hline\hline
\textbf{Beamline}                     & \textbf{I05}     & \textbf{URANOS}    & \textbf{ADRESS} \\
\hline
Sample                                & Tl2504           & Tl2504             &  Tl2201~\cite{kramerPRB2019,horio_three-dimensional_2018}\\
$h\nu$ (eV)                           & 60               & 100                & 428 \\
Spot size (\si{\micro\meter^2})       & $50 \times 50$   & $60 \times 150$    & $10 \times 74$\\
Polarization                          & $\pi$            & $\pi$              & $\sigma$ \\
$E_{res}$ (meV)                       & 10               & 20                 & 90 \\
$T$~(K)                                 & 200              & 200                & 20 \\
%\hline\hline
\end{tabular}
\end{ruledtabular}
\label{tab:ARPES}
\end{table}

\noindent\textbf{Modeling of magnon dispersion:}
The fitting of magnon excitations is based on a Heisenberg Hamiltonian derived from the Hubbard model. The magnon dispersion is parametrized by Hubbard repulsion $U$ and nearest-neighbor hopping $t$, as well as high-order hopping terms, i.e., next- ($t^{\prime}$) and next-next ($t^{\prime\prime}$) nearest-neighbor hopping. $\hbar\omega_k= Z_c(U,t,t^\prime,t^{\prime\prime})\epsilon_k(U,t,,t^\prime,t^{\prime\prime})$, where $Z_c$ is a momentum-dependent renormalization factor and $\epsilon_k(U,t,,t^\prime,t^{\prime\prime})=\sqrt{A_k^2-B_k^2}$ is a bare dispersion with $A_k$ and $B_k$ factors determined by the fitting parameters, as described in~\cite{delannoy_low-energy_2009,dalla_piazza_unified_2012,ivashko_damped_2017}. The ratio is set to $t^{\prime\prime}/t^\prime=-0.5$, based on experimental photoemission results and DFT calculations~\cite{yoshida_systematic_2006,matt_direct_2018,sakakibara_two-orbital_2010}. The fitting parameters are obtained by minimizing $\chi^2$. Extended fitting results for all cuprate materials presented in Fig.~\ref{Fig4} are included in SI, Table~S1.
\\

\noindent\textbf{Acknowledgments}\\
We acknowledge H.~Eisaki for providing Tl2504 single crystals. We acknowledge C.~C.~Tam for sharing data on Tl2201. I.B. acknowledges support from the Swiss Confederation through the Government Excellence Scholarship ESKAS-Nr:~2022.0001. I.B. and K.v.A. thank the Forschungskredit of the University of Zurich, grant no. FK-23-113 and FK-21-105. K.v.A. and J.C. acknowledge support from the Swiss National Science Foundation (SNSF) through Grant Numbers BSSGI0-155873, and 200021-188564. K.v.A. thanks for the support from the FAN Research Talent Development Fund - UZH Alumni. Q.W. and Y.C. thank the support from the Research Grants Council of Hong Kong (ECS No. 24306223) and the Guangdong Provincial Quantum Science Strategic Initiative (GDZX2401012). W.P. was supported by the SNSF under project no. 200021\_185037. Y.S. thanks the Chalmers Area of Advances-Materials Science and the Swedish Research Council (VR) with a starting Grant (Dnr. 2017-05078) for funding. This publication was partially developed under the provision of the Polish Ministry and Higher Education project "Support for research and development with the use of research infra-structure of the National Synchrotron Radiation Center SOLARIS” under contract no 1/SOL/2021/2. Sample preparations and characterizations were done using glove box and MPMS at the CROSS Laboratory. We acknowledge Diamond Light Source for access to beamline I05 (proposal number SI32147) that contributed to the results presented here. We acknowledge the European Synchrotron Radiation Facility (ESRF) for provision of synchrotron radiation facilities under proposal number HC-4422. \\

\noindent\textbf{Author contributions}\\
M.I. grew the Tl2201 and Tl2504 single crystals. C.L., I.B., J.K., W.P., O.K.F., M.R., N.O., M.D.W and T.K.K. carried out the ARPES experiments. I.B., L.M., X.H., O.G., Q.W., K.v.A, D.B, N.B., J.C. carried out the RIXS experiment. Q.W., I.B., X.H. and Y.C. developed the magnon fitting procedure. I.B. carried out the data analysis. I.B. and J.C. wrote the manuscript with input from all the other authors.\\

\noindent\textbf{Data availability} \\
The data that support the findings of this article are openly available: https://doi.org/10.5281/zenodo.17642238.\\

\noindent\textbf{Code availability} \\
The data analysis code supporting the findings of this study is available from the corresponding author upon reasonable request. \\

\noindent\textbf{Competing interests} \\
The authors declare no competing interests.\\

\bibliography{Tl2504_bib}

\end{document}

\begin{center}
    \Large\textbf{Supplementary Information: Magnetic Excitations of a Half-Filled Tl-based Cuprate}\\
    \vspace{0.4cm}
    I.~Bia\l{}o \textit{et al.}
\end{center}

\vspace{0.3cm}

\section{Crystal structure}
\begin{figure*}[h]
    \centering
    \includegraphics[width=\linewidth]{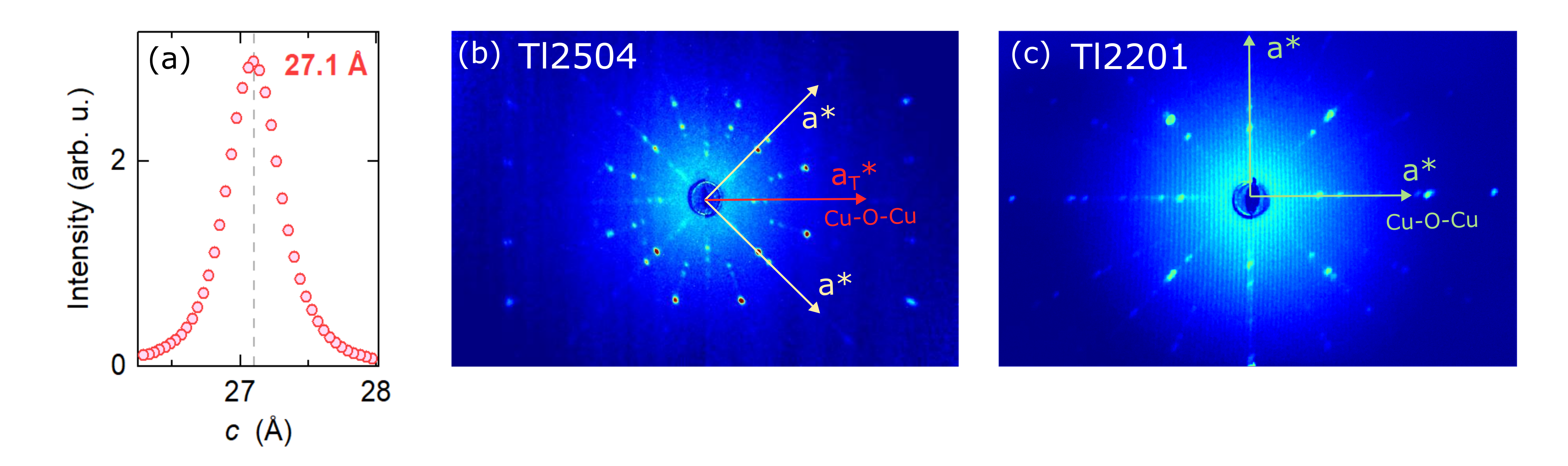}
    \caption{\textbf{Crystallographic characteristics of the Tl-based cuprates.} (a) X-ray diffraction measurements of the \textit{c}-axis parameter performed at the ID32 beamline in ESRF synchrotron for Tl$_2$Ba$_5$Cu$_4$O$_{x}$ (Tl2504). The size of a unit cell confirms the bilayer structure of Tl2504 characterized before in Ref.~\cite{Hasegawa1996}. (b,c) X-ray Laue patterns of Tl2504 (b) and Tl$_2$Ba$_2$CuO$_{6+\delta}$ (Tl2201) single crystals. The CuO$_2$ plane of Tl2504 crystal structure is 45$^\circ$ rotated along $c$-axis in reference to a unit cell of Tl2201. Therefore, for Tl2504, We adopt a tetragonal reference unit cell with $a_{\mathrm{T}}$ to simplify the description of magnetic excitations in comparison with other cuprates. }
    \label{Laue_fig}
\end{figure*}

\section{Superconducting properties of Tl2201}
\begin{figure*}[h]
    \centering
    \includegraphics{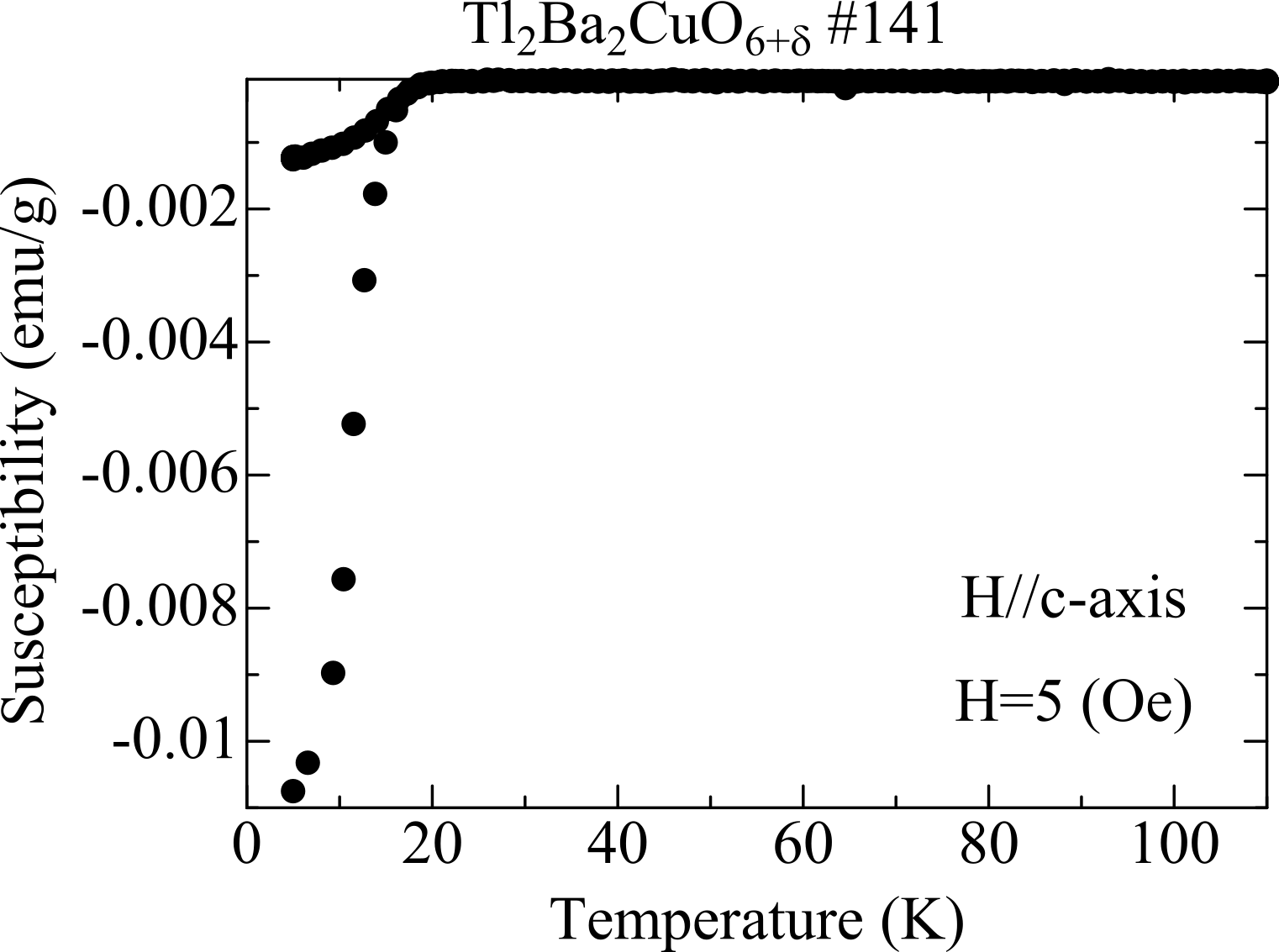}
    \caption{\textbf{Critical temperature of Tl2201.} Magnetization measurements of Tl2201 with $T_c=20$~K. Measurements conditions are indicated in the figure. Data for the sample presented in the Fig.~1(d) of the main text, adapted from~\cite{horio_three-dimensional_2018}.}
    \label{FigS2}
\end{figure*}

\section{Multimagnon excitations}
The low-energy (as defined in the main text) part of the Tl2504 RIXS spectrum was analyzed using a four-component model. The elastic contribution, phonon, and single-magnon excitations are fitted with the resolution line shape, a Gaussian profile. The remaining low-energy spectral weight forms a broad, continuous distribution of spectral weight in the $180 - 500$~meV energy range. It is fitted with a wider Gaussian profile, reaching about $300$~meV of FWHM (Full Width at Half Maximum). This broad spectral weight is not uniform across the Brillouin zone. It is most intense near the $(0.5,0)$ point, consistent with previous reports. Different interpretations, such as  magnon–magnon interactions~\cite{ChristensenPNAS2007,powalski_mutually_2018} and magnon fractionalization~\cite{martinelli_fractional_2022,dalla_piazza_fractional_2015}, have been put forward.
\vspace{2cm}

%This scattering continuum is widely discussed in the cuprate literature~\cite{MartinelliPRX2022,ChristensenPNAS2007,HeadingsPRL10,dalla_piazza_fractional_2015}. It might be interpreted as arising from magnon–magnon interactions leading to renormalized multimagnon excitations. Because the coupling between multiple spin flips depends on both energy and momentum, the corresponding spectral intensity is not uniform across the Brillouin zone. Its spectral weight increases near the $(0.5,0)$ point in the Brillouin zone, which has been suggested to be a result of magnon fractionalization~\cite{MartinelliPRX2022,dalla_piazza_fractional_2015}.}

\section{Comparison of magnon dispersion in cuprates}
\begin{figure*}[h]
    \centering
    \includegraphics[scale=1]{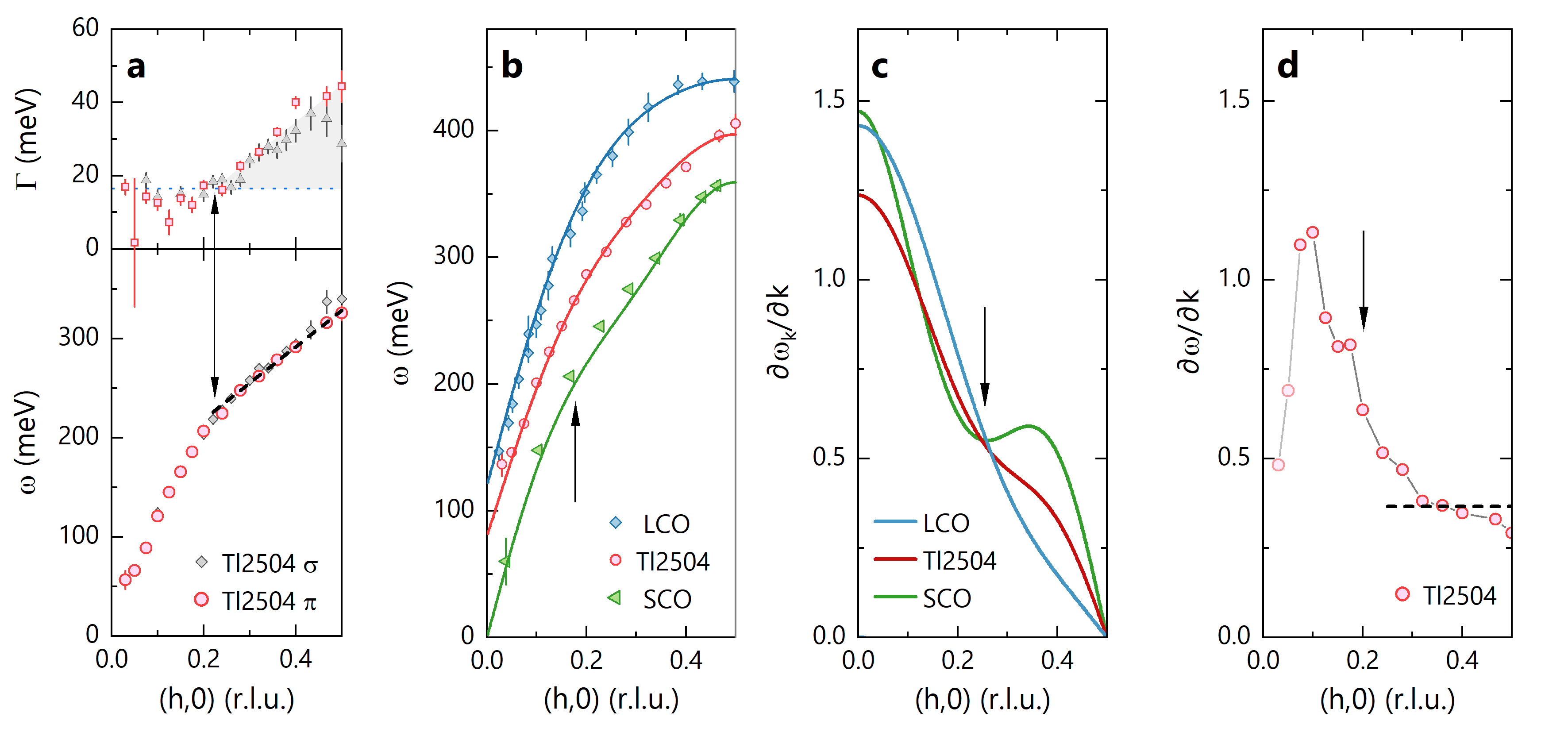}
    \caption{\textbf{Magnon dispersion and its derivative} (a) Single-magnon dispersion and its inverse lifetime extracted from RIXS measurements performed with $\sigma$ and $\pi$ polarized light. The black arrow indicates simultaneous change of magnon energy and inverse lifetime. The gray shaded area presents the single-magnon broadening near the $(0.5,0)$ point. A blue dotted line indicates the experimental energy resolution. A black dashed line is a linear fit to the selected range of the magnon dispersion. (b) Comparison of magnon dispersion along the $(h,0)$ direction for selected cuprate systems. Symbols represent experimental data. Solid lines are corresponding fits using the Hubbard model including renormalization factor (see the main text). Datasets for Tl2504 and LCO are shifted by a constant in energy scale, correspondingly by $0.8$~eV and $0.12$~eV. Data for LCO (SCO) has been adapted from~\cite{headings_anomalous_2010} (\cite{wang_magnon_2024}). (c-d) First derivative of: (c) the modeled magnon dispersion from (b), (d) the experimental data of Tl2504 taken with $\pi$-polarized light. (d) The dashed line indicates the slope of the magnon dispersion extracted from linear fit in (a).}
    \label{FigS3}
\end{figure*}

\newpage

\section{Modeling of magnon dispersion}

\begin{figure*}[h]
    \centering
    \includegraphics[scale=1]{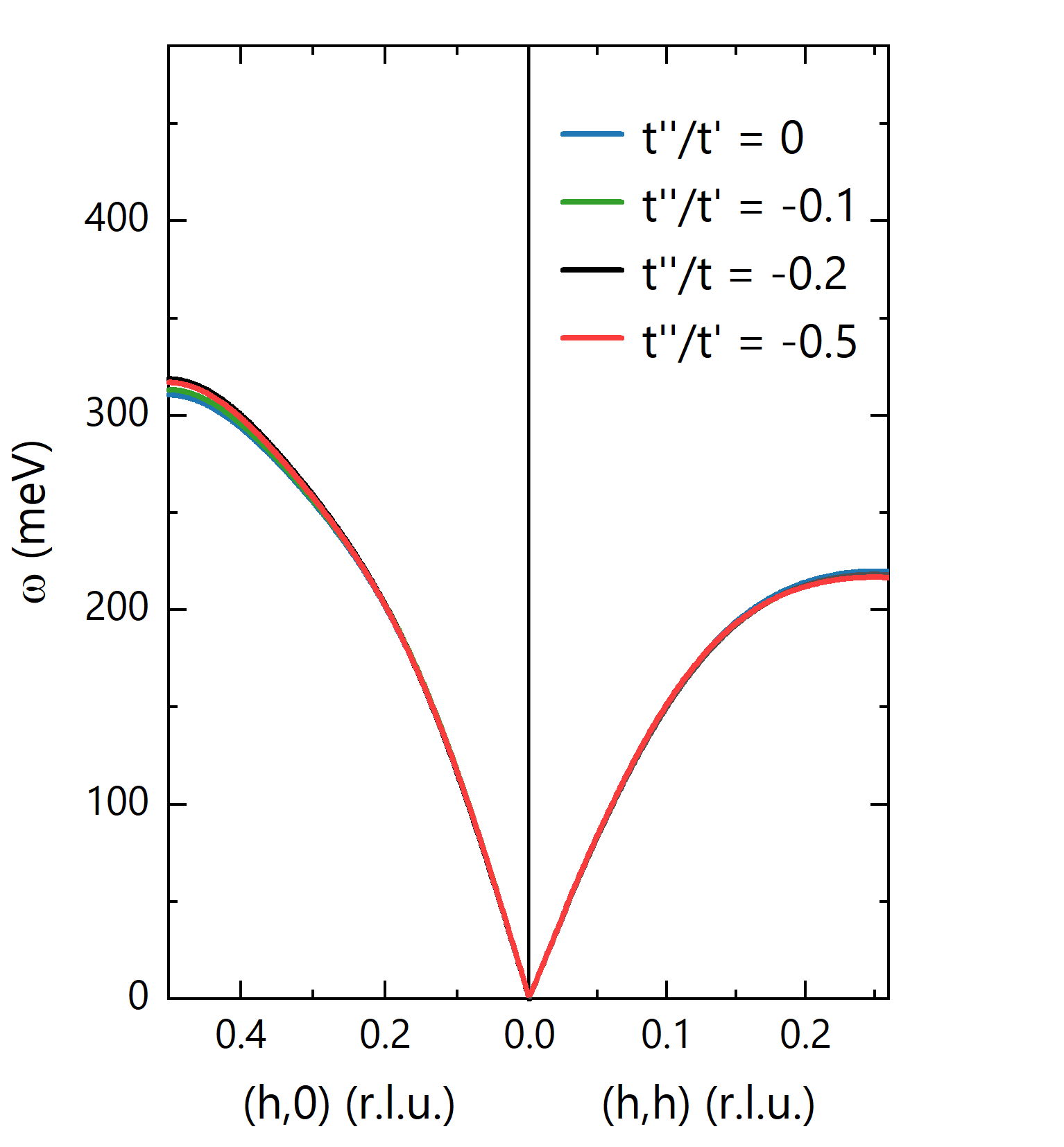}
    \caption{\textbf{The fitting model for varied $t''/t'$ ratio.} Values of $t''/t'$. The results of fitting of the Tl2504 magnon dispersion performed for a different fixed $t''/t'$ ratio. The result of the fitting is very weekly affected by the $t''/t'$ value.}
    \label{FigS4}
\end{figure*}

\begin{table}[h]
\centering
\setlength{\tabcolsep}{12pt} % default is usually 6pt
\renewcommand{\arraystretch}{1.3} % increase row height for readability
\begin{tabular}{lccccc}
\hline
\textbf{Sample} & \textbf{$E_{\mathrm{ZB}}$} & \textbf{$U/t$} & \textbf{$t^{\prime}/t$} & \textbf{$t^{\prime\prime}/t$} & \textbf{$Z_{c,ave}$} \\
\hline
LCO         & 39.64  & 9.79  & -0.44 & 0.22  & 1.53 \\
SCOC        & 43.07  & 9.33  & -0.44 & 0.22  & 1.56 \\
Bi2201      & 76.59  & 8.24  & -0.44 & 0.22  & 1.70 \\
Tl2504      & 100.18 & 6.91  & -0.38 & 0.19  & 1.53 \\
CCO         & 150.66 & 7.00  & -0.43 & 0.22  & 1.97 \\
SCO         & 165.29 & 6.36  & -0.41 & 0.20  & 2.00 \\
\hline
\end{tabular}
\caption{\textbf{The optimized Hubbard-Heisenberg model parameters for cuprate materials.} Summary of the fitting results of magnon dispersions performed using the Hubbard model with momentum dependent renormalization factor $Z_c$, as described in~\cite{delannoy_low-energy_2009,dalla_piazza_unified_2012,ivashko_damped_2017}. $Z_{c,ave}$ stays for an average value of the renormalization factor within the fitting range. Experimental magnon dispersions used for fitting come from Tl2504 (this work), La$_2$CuO$_4$~\cite{headings_anomalous_2010}, Sr$_2$CuO$_2$Cl$_2$ (SCOC)~\cite{plumb_high-energy_2014}, Bi$_2$Sr$_{0.9}$La$_{1.1}$CuO$_6$ (Bi2201)~\cite{peng_influence_2017}, CaCuO$_2$ (CCO)~\cite{martinelli_fractional_2022} and SrCuO$_2$ (SCO)~\cite{wang_magnon_2024}. }
\label{tab:model_params}
\end{table}

\newpage
\bibliography{Tl2504_bib}